\documentclass[twocolumn,aps,pre]{revtex4}

\usepackage{xcolor}
\usepackage{graphicx}
\usepackage{dcolumn}
\usepackage{amsmath}
\usepackage{latexsym}
\usepackage{verbatim}

\begin {document}

\title{Stochastic evolution of interacting directed strings}
\author{S. S. Manna\footnote[1]{subhrangshu.manna@gmail.com}}
\affiliation{B-1/16 East Enclave Housing, 02 Biswa Bangla Sarani, New Town, Kolkata 700163, India}
\author{Deepak Dhar\footnote[2]{deepakdhar1951@gmail.com}}
\affiliation{ International Center for Theoretical Sciences, Shivakote, Hesaraghatta Hobli, Bengaluru 560089, India}

\begin{abstract}
      We study the stochastic dynamics of a system of $N$ polymer rings on an infinite cylinder with only excluded volume interaction, using simulations  and also analytically. The strings can diffuse in the transverse  direction  by local Rouse-like moves where an up-kink changes to a down -kink, or vice versa.  The motion is biased if the rates of flip are not equal. In the initial state, we start with  a step initial condition: all the sites with a positive $y$ coordinate are empty, and the sites with $y \leq 0$ are occupied with a uniform density.  We study the scaling form of  the density profile in the hydrodynamic limit. The scaling form is a function of $y/\sqrt{t}$ in the unbiased case, and a function of $y/t$ in the biased case. In the unbiased case, in the steady state on a torus, the marginal probability distribution of the positions of the strings on a line with fixed $x$-coordinate is exactly same as that in  the ground state of quantum-mechanical  spin-less fermions on a line.  The biased problem is equivalent to a growing surface in $2+1$ dimension, which is expected to be  in the Kardar-Parisi-Zhang universality class. 
\end{abstract}

\maketitle

\section {INTRODUCTION}

\begin{figure}
\includegraphics[width=7cm]{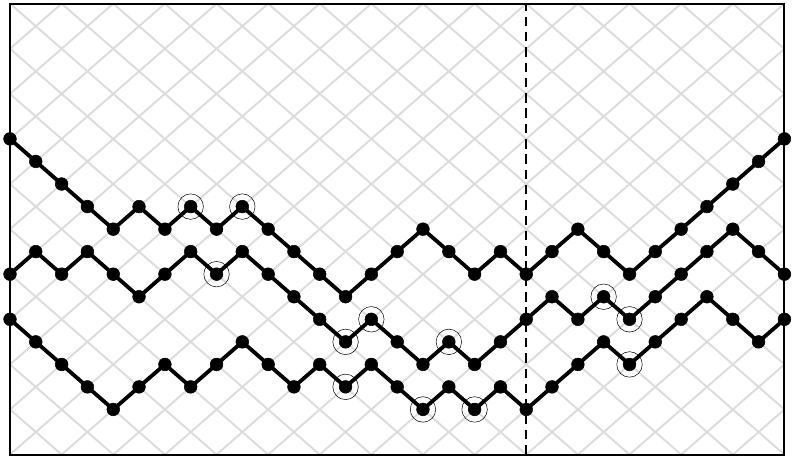}
\caption{An example of a configuration of $3$ strings on a cylinder. The 12 circles mark the non-flippable kink sites.  The dotted vertical marks a particular $x$, equivalently a particular the quantum mechanical time, in the equivalent quantum mechanical fermions on a line problem, discussed later.  The positions of the monomers at this line  are  $Y_1, Y_2, Y_3$.}
\label{figure01}
\end{figure}

    This work deals with the hydrodynamical description of a system of interacting polymers in solution.  Polymer dynamics  of a set of flexible linear polymers in a dense environment of other similar polymers has been studied for  a  long time.  De Gennes, and Doi and Edwards studied the `polymer chain in a tube’ model which considers a long polymeric molecule constrained to move within some effective ‘tube’ due to the  excluded volume interaction with the other chain molecules in a  snake-like motion called reptation \cite{Doi, Gennes, Gennes1}. These models  deal  with diffusion of a single polymer chain  in an environment provided by other chains. In this paper, we will study  diffusion of chains interacting with each other by excluded volume interaction, with  all  chains is treated on equal footing.  We will study this problem in the  hydrodynamical limit.  The subject of hydrodynamical limit of interacting point particles has attracted  much interest  in recent years \cite{ Kipnis, scaling_limit2}.  When the interacting constituents are extended  deformable objects, such as strings, much less is known. In general, the fluctuations of the density of polymer chains will be anisotropic, with displacements along  the chain length having a different behavior that transverse fluctuations. Here we will study the simpler case of fully directed strings in two dimensions, where only transverse fluctuations are possible [Fig. \ref{figure01}]. 
    
    Our model is  also equivalent to a model of a growing $2$-dimensional surface, and thus this is also a contribution to study of the Kardar-Parisi-Zhang equation in $2+1$ dimensions. While the behavior of the KPZ equation in $1 + 1$ dimensions is well understood, its behavior in higher dimensions is not so \cite{kpz0,kpz1, kpz2,kpz3, kpz4}.     We  will discuss the model, both with and without a bias field.  Numerical simulations to determine the critical exponents in  the $2+1$ dimensions will be discussed in a forthcoming publication.  The model we discuss is also of interest, as  the steady state of the model  is related  to several well-studied models in equilibrium statistical mechanics   The steady state of the unbiased model can be described in terms of   a free fermion model in $1+1$ dimensions, which is a paradigmatic model in equilibrium statistical physics, with a long history, starting with the ice model \cite{ice}, the Ising model \cite{ising}, and the dimer model on the hexagonal lattice \cite{kasteleyn}, and also  the trimer model on the kagome lattice \cite{trimers}.

 For the unbiased problem, in  the  hydrodynamic limit, the evolution of  small-amplitude large-wavelength  density perturbations over a background uniform density $\rho$ involves the bulk-diffusion constant $D(\rho)$, that depends on the density $\rho$. The hydrodynamic evolution  equation  turns out to be what is known as  the porous medium equation  in literature \cite{porous_medium_1, porous_medium_2,porous_medium_3}.    For the one dimensional case of hard core particles, it is known that the bulk diffusion constant is finite, and positive, and in fact independent of density \cite {Arratia, Kipnis}. In our problem, 
we are not able to determine this diffusion constant from first principles, and determine it experimentally  from simulations.  Surprisingly, we find that this diffusion constant is an {\it increasing} function of density,  even though the only interactions in the problem are the  hard-core interactions. We  study the evolution of the system on a cylinder, from step initial conditions in density, and determine the asymptotic profile in the hydrodynamic limit in terms of the function $D(\rho)$ determined from simulations. 
      
      The steady state of unbiased motion of $N$ interacting strings on a  finite torus,  can be characterized exactly.   Our model evolves under classical stochastic dynamics,  but  it is equivalent to the  time-history of evolution of the quantum mechanical system of free fermions:  If  we look at  a randomly picked configuration  in the steady state,  the marginal joint probability distribution of  the positions $\{R_i\}$   where the $N$ strings intersect any one  line $x =$ constant,  away from the boundaries,  is  exactly the same  is in the quantum -mechanical wave function of the ground state of free fermions. We also  determine exactly the entropy per monomer in the steady state as a function of the density. 
      
      For the case of biased motion,  for a given strength of the bias, we also consider the steady state of the system on a torus (periodic boundary conditions). If the rates of flips of up-kinks and down-kinks are $\lambda$ and $1$, with $\lambda <1$, the bias produces a non-zero drift velocity  in the transverse direction. In the limit of large systems, this mean velocity  is a function of the density of the strings, and the bias strength , and denoted by  $v(\lambda, \rho)$.   Our numerical studies suggest that $v(\lambda, \rho)$ is a strictly linear function : $ v(\lambda,\rho) = (1 -\lambda) v^*(\rho)$.    This function $v^*(\rho)$ is also determined  only from the simulation data.  Using  numerically determined function $v^*(\rho)$, and mass conservation, we can determine  the scaling function of density  profile $H(y/t)$ for the step  initial condition  exactly in terms of  the function $v^*(\rho)$ by a simple equation, that does not involve the solution of any differential equation.  We find that the  scaling function $H(u)$ is very well approximated by a simple algebraic function. 

The plan of this paper is as follows: in Section II, we define the model, and discuss its equivalence to growing surfaces in $(2+1) $ dimensions. In Section III, we discuss the  hydrodynamic equation that  determines the evolution of the density  in the unbiased case.  We show that the scaling limit of the density profile can be determined in terms of a phenomenological density-dependent transverse diffusion constant.  We argue that this diffusion constant varies as $\rho^2$ for small densities $\rho$.  In Section IV, we discuss  the calculation of stationary measure in the steady  state of the system. In the transfer matrix formalism, this  becomes equivalent to the calculation of ground state spinless fermions in one dimension.  This also determines the entropy per site in the steady state as a function of the density. We present a heuristic argument for the quadratic variation of diffusion constant as a function of density by estimating the entropic force for reducing density gradients. In Section V, we discuss  the biased case.  We first show that the steady state measure for biased diffusion is nontrivial. Then present numerical data for the mean velocity as a function of the bias and density. We use this to study the evolution of  the density profile with time, starting with the step initial condition.   We show that in the scaling limit, the  density profile is a function only of $y/t$, and this scaling solution can be determined in terms of the density-dependent drift velocity. Section VI contains a summary and some concluding remarks. 

\begin{figure}[t]
\includegraphics[width=6.0cm]{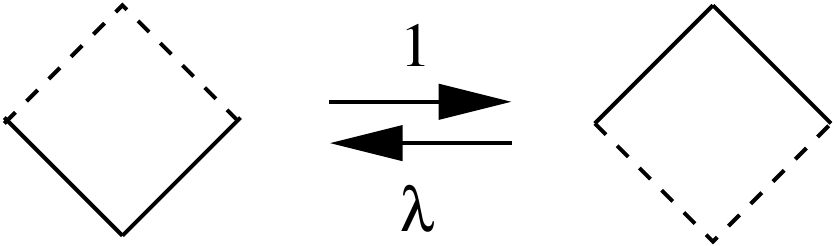}
\caption{The possible moves for transverse fluctuations of a string, where an up kink can become a down kink, 
or vice versa. The move is allowed only if the target site is originally empty.}
\label{figure02}
\end{figure}

\section {Definition of the model}

      We start with  an oriented square lattice of size 
   $L_x \times L_y$, with $L_x$ and $L_y$ both even,  with periodic boundary condition along the $x$ axis. A string is a set of steps, with $\Delta x =1, \Delta y = \pm 1$, which comes back to the starting point after $L_x$ steps, and hence is a ring.   We place  $N$ rings parallel to each other. Initially, each string comprises  of an alternate sequence of up and down kinks. In the actual computer implementation, 
   we used for the coordinate system a regular square grid $(x,y)$, with $1 \leq x \leq L_x$, $1 \leq y \leq L_y$, 
   but the lattice sites are only those  with $x+y$ even. Then, the number of allowed sites in a lattice 
   of size $(L_x, L_y)$, is $L_x L_y /2$.
   An example of a configuration of $3$ strings placed on cylinder with $L_x= 30$ and $L_y = 20$ is shown in Fig. \ref{figure01}.

     The system undergoes a Markovian evolution in continuous time.  During the dynamical evolution of the string system a single microscopic move consists of converting  a down kink to an up kink which we refer as the `upward' move. A move is allowed only if it does not violate the excluded volume constraint, i.e., the overlap    of monomers is forbidden [Fig. \ref{figure02}].  If allowed, it occurs with rate $1$. The    opposite, i.e., from an up kink to a down kink and call it as the `downward' move, occurs with rate    $\lambda$. If $\lambda =1$, this is called unbiased diffusion.    If $\lambda <1$, it is called biased diffusion, The case $\lambda=0$, where the downward moves are not allowed, will be called fully biased. 
   
\begin{figure}
\includegraphics[width =5cm]{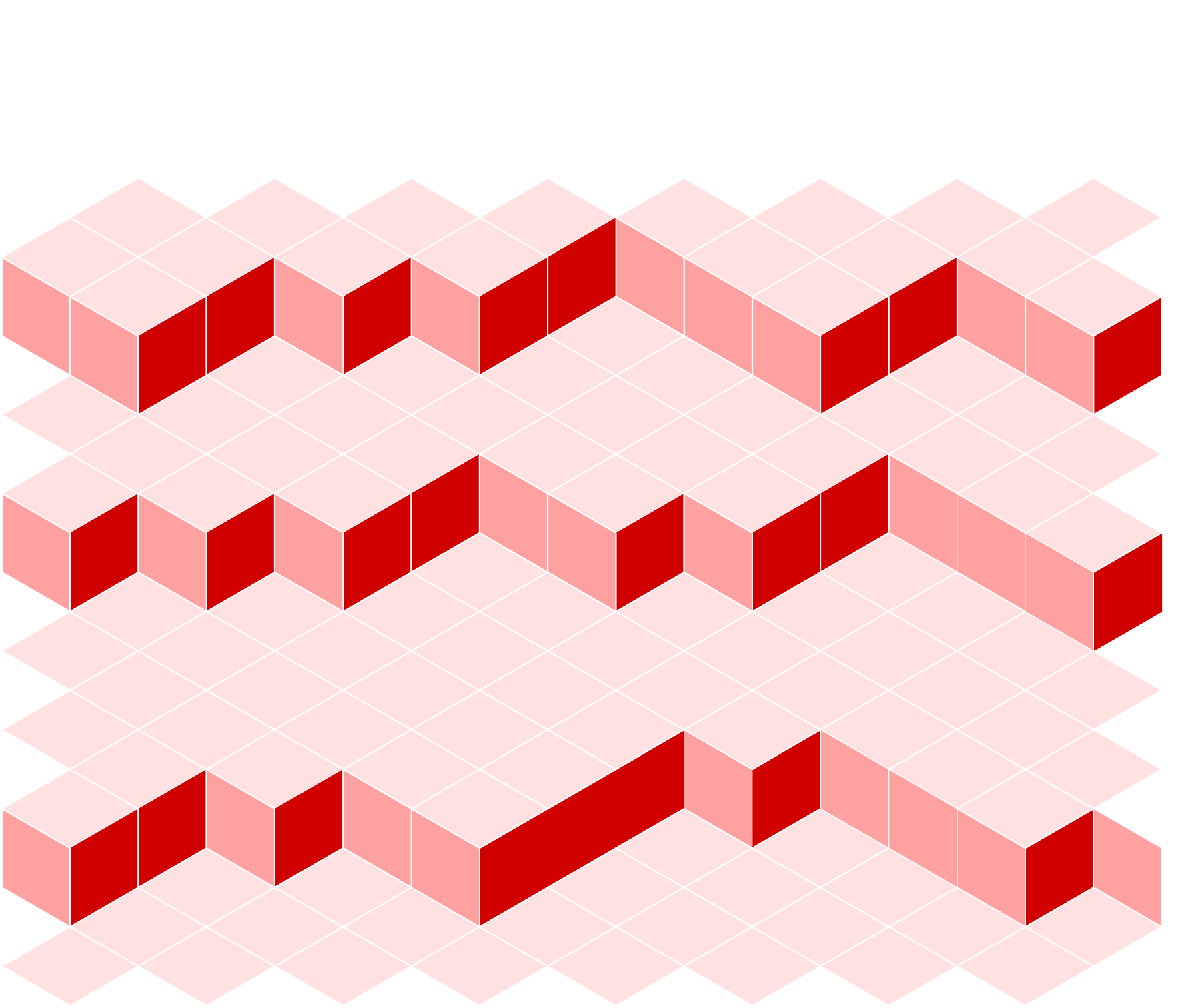}
\caption{A schematic representation of a stepped surface. The $z$ coordinate of the surface is a 
   non-decreasing function of $x$ and $y$ coordinates, and maximum difference between height at a 
   site $(x,y)$ and a neighboring site is at most $1$.}
\label{figure03}
\end{figure}

      The biased diffusion of the set of strings may also be viewed as the  growth of a  surface in $2+1$ dimensions. The configurations of strings on the infinite cylinder are in one-to-one correspondence with configurations of a stepped Solid-on-solid (SOS) surface in $2+1$ dimensions (Fig. \ref{figure03}) . 
           
      In the SOS model, the occupied region by the atoms constituting the 3-dimensional solid is separated from the the unoccupied region by a surface $z=  h(x,y)$, where a  site $(x,y,z) $  of a cubic lattice is occupied iff $z \leq h(x,y)$. We will further assume that $h(x,y)$ is a non-decreasing function of the coordinates $x$ and $y$, and the values of heights at nearest neighbor sites can differ by at most $1$. Then the surface is made of steps of height 1 [Fig. \ref{figure03}], and the boundaries of regions of constant height on the surface, projected on to the $(x,y)$ plane are the directed strings in our model.  Then, a kink motion in the biased diffusion of strings corresponds to a particle getting attached or detached from the surface at a corner, keeping the SOS constraint, with probability of attachment greater than the probability of detachment. This model has been studied earlier by many authors. See, for example, \cite {Forrest, Forrest1, Tang, borodin}.    This  gives us a computationally efficient  way to simulate such  growing surfaces.   Its  critical behavior would be, by definition,  in the Kardar-Parisi-Zhang (KPZ) universality class in $2+1$ dimensions. The Monte Carlo studies of correlations of fluctuations, and the estimation of the KPZ exponents will be discussed in a future publication. 
            
\section {Unbiased diffusion}

We have  studied  the evolution in the absence of bias using the Monte Carlo simulations in both the non-stationary as well as the stationary states. In the non-stationary state we studied    how a density discontinuity  in the initial state relaxes under the flipping dynamics.  If we   put the strings on a finite torus, with periodic boundary conditions in both $x$- and $y$-directions, then at long times, we get a stationary state.  

\begin{figure}[t]
\begin {tabular}{cc}
\includegraphics[width=4.0cm]{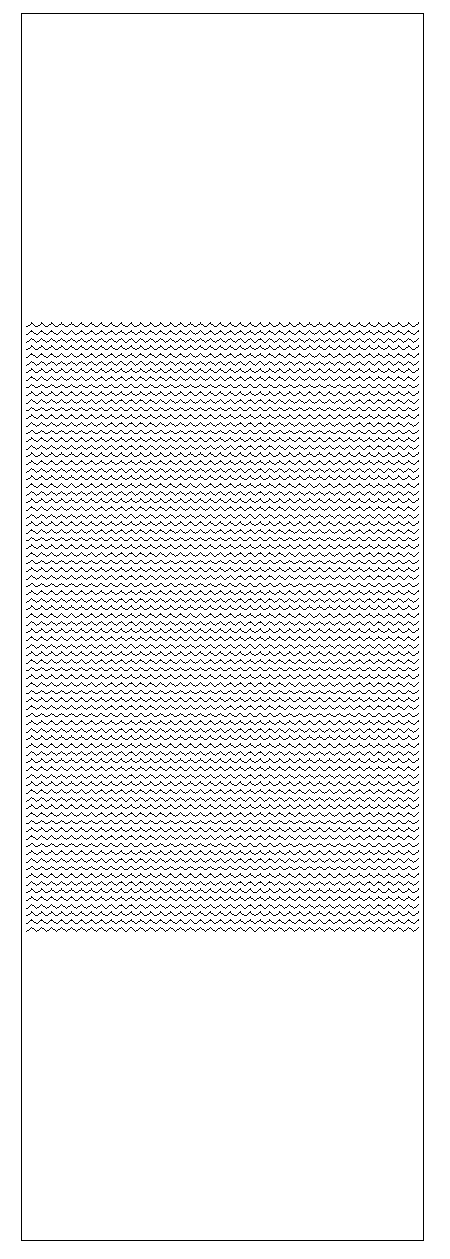} &
\includegraphics[width=4.0cm]{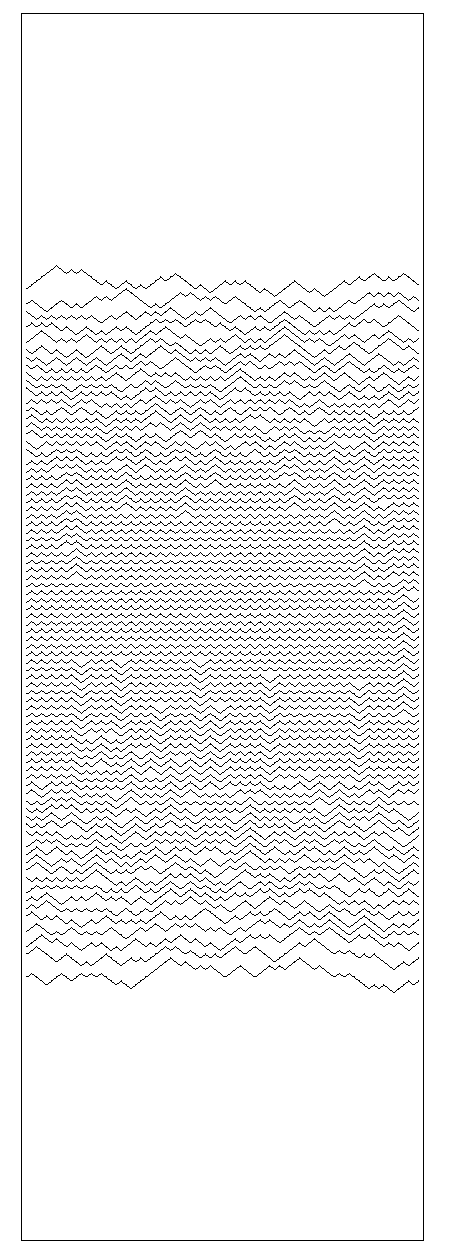}
\end {tabular}
\caption{Unbiased diffusion:
(a) The initial  set of $N$ = 80 close-packed  strings on the oriented square
lattice drawn on the infinite cylinder of width $L_x = 80$.
(b) A typical non-equilibrium configuration of the same set of strings when 
evolved $t = 100$ time steps under the unbiased evolution process.
}
\label {figure04}
\end{figure}

\begin{figure}
\includegraphics[width=8.5cm]{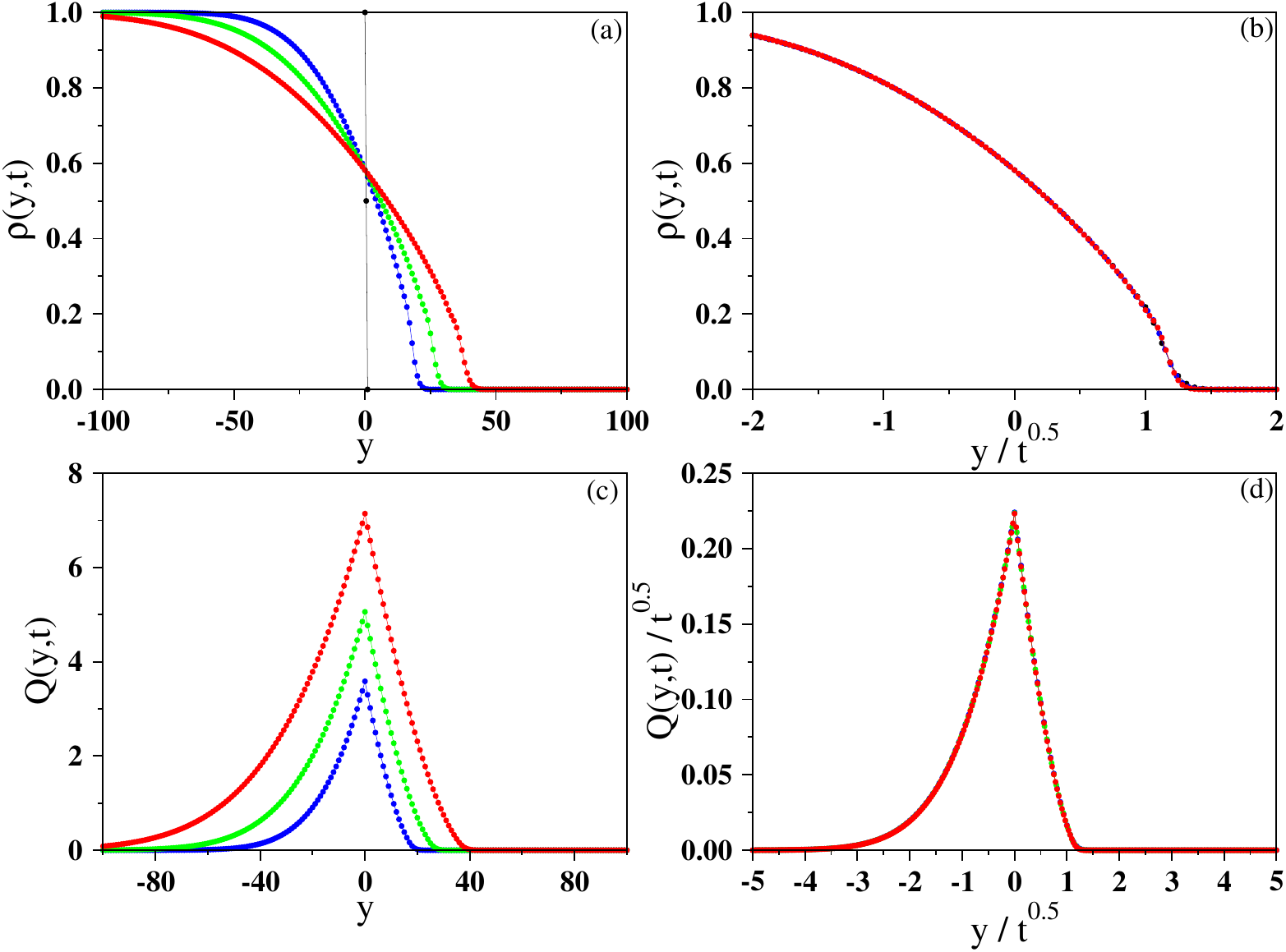}
\caption{   (a) The average density $\rho(y,t)$ profile of monomers of a dense set of  strings covering   the lower half on an infinite cylinder of width $L_x$ = 128 units, for time instants 0 (black), 256 (blue), 512 (green), and 1024 (red).
   (b) Scaling plot of $\rho(y,t)$ against $y / \sqrt{t}$. (c) The integrated current density $Q(y,t)$ has been plotted against the vertical coordinate $y$ for a system of 
    unbiased strings on a lattice of lateral width $L_x$ = 256 at time instants: $t$ = 256 (black), 512 (blue) and 1024 (red).
(d) The scaling of the current densities. The scaled plot $Q(y,t) / \sqrt{t}$ against $y / \sqrt{t}$ 
    exhibits an excellent data collapse for all three times $t$.  }
\label{figure05}
\end{figure} 

      Initially, each string of length $L_x$ is placed fully stretched along  the $x$-axis and consists of an alternate sequence of top-right and down-right bonds. We start by densely packing  $N$ strings in a finite region of width $2N$ of the cylinder (each zig-zag string occupies a strip of width $2$ ).   A typical string configuration in this case is shown in Fig. \ref{figure04}(a). In the computer implementation of the continuous -time Markov process, we use the random sequential update:  In a single step of the Monte Carlo simulation, a monomer out of the $N L_x$ monomers possible is picked at random, and moved if the target site is not already occupied by another string, and the clock time is increased by $1/(NL_x)$. We try to move the randomly selected kink irrespective of whether it is an up or a down kink. A typical resulting configuration after many such moves is  shown in Fig. \ref{figure04}(b). 
   
      As is seen from Fig. \ref{figure04}(b), the monomer density becomes smoother as time 
   proceeds. Since there is translational invariance along the $x$ axis, the average monomer density explicitly depends only on the $y$ coordinate and time. We have measured the average monomer density $\rho(y,t)$, i.e., the fraction of  lattice sites on a horizontal line with the vertical coordinate $y$ that are occupied. On an infinite cylinder of circumference $L_x$, at time $t=0$, the  monomer density profile is a step function.  
 
      If the initial density is not uniform,  diffusion  will tend to reduce the density gradients. In a hydrodynamical theory, in  the linear response regime,  the instantaneous current density $j(y,t)$ is  expected to be proportional to the local density gradient:
\begin {equation}
j(y,t) = - D(\rho) \frac {\partial \rho}{\partial y}
\label{eq:fick_law}
\end {equation}
   where $D(\rho)$ is the position dependent local diffusion constant and is an explicit function of 
   the density $\rho(y,t)$ in general. The rate of change of the density $\rho$ with time 
   is given by the divergence of the current density by the continuity equation
\begin {equation}
\frac {\partial \rho}{\partial t} = \frac {\partial}{\partial y}[D(\rho) \frac{\partial \rho}{\partial y}].
\label{eq:unscaled_diffusion_equation}
\end {equation}

This equation is known as the porous medium equation in literature, and has been studied a lot, as it is encountered in many contexts.  Several properties are known, specially for the case when $D(\rho)  = \rho^m$, with $m  >1$, See for example, the reviews \cite{porous_medium_1, porous_medium_2, porous_medium_3}. In particular, for a point source, explicit  analytical  solutions have been obtained in many cases using the scaling property. See, for example \cite{king}.

This equation is invariant under $y \rightarrow ay$, $t \rightarrow a^2t$. We find that at later times    the density profiles can be collapsed on a scaling curve $\rho(y,t)  =  \rho(y/\sqrt{t}).$  We plot the density $\rho(y,t)$ against $y$ for $L_x = 128$ at different time instants $t$ = 0, 256,    512, and 1024 in Fig. \ref{figure05}(a).
In Fig. \ref{figure05}(b), we plot the same data using $y/\sqrt{t}$ as the scaling variable.  We find an excellent collapse. In particular, all the curves in Fig. \ref{figure05}(a) have a common point $(\rho_c,0)$, where $\rho_c \approx 0.5888$, which is  consistent with the scaling property, where for all times, $y=0$ has the same scaled value of $y/\sqrt{t}$.       We define the integrated current density $Q(y,t)$ as the  number of particles crossing the line with y-coordinate $y$ up to the time  $t$ per unit length.  

In Fig. \ref{figure05}(c), we plot the numerically calculated integrated current density $Q(y,t)$ for three different values of $t$.  Fig. \ref{figure05}(d) shows the scaling collapse, when $Q(y,t)/\sqrt{t}$ is plotted against $y/\sqrt{t}$. 

   In terms of the scaled variable $u=y/\sqrt{t}$ the equation takes the form:
\begin {equation}
-u\frac{d \rho}{du} = \frac{d}{du}[D(\rho)\frac{d\rho}{du}].
\label{eq:unbiased_diffusion}
\end{equation}
   The two functions $D(\rho)$ and $\rho(u)$ are dependent on each other. If one is known, the 
   other can be  determined.

\begin{figure}[t]
\includegraphics[width=6.0cm]{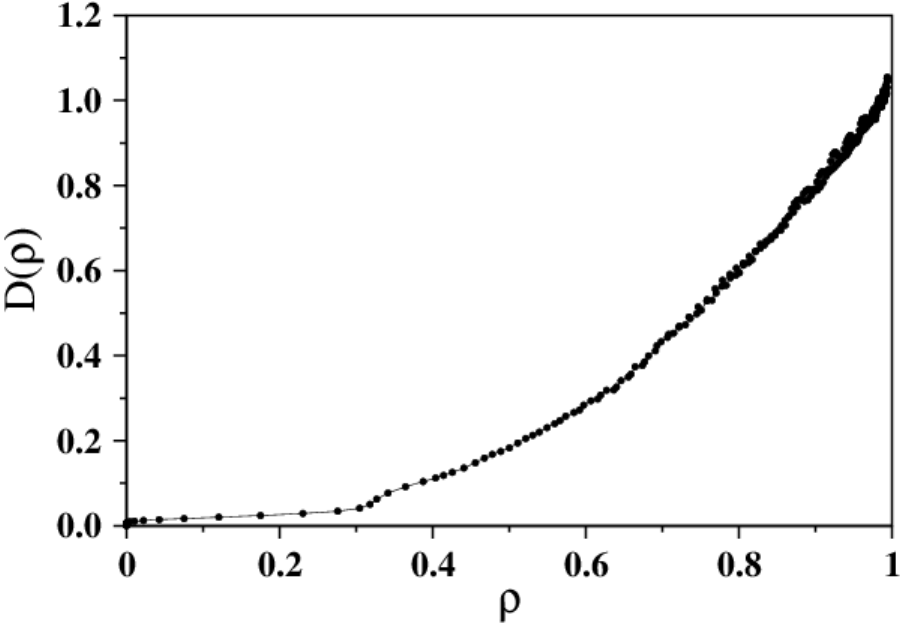}
\caption{Unbiased diffusion:
The diffusion constant D($\rho$) has been plotted against the monomer 
density $\rho$. The string system starts from a half-filled infinite cylinder ($y \le 0$) and is empty 
above, with a lateral size $L_x$ = 512, and evolves till $t$ = 1024 time steps.
}
\label {figure06}
\end{figure}

The diffusion constant $D(\rho)$ is numerically determined  by taking the ratio of the instantaneous current density    $j(y,t)= \frac{\partial}{\partial t} Q(y,t)$ and the 
   density gradient $\partial \rho/\partial y$, for different values of $y$ and $t$.  We plot in Fig. \ref{figure06} the $D(\rho)$ 
against $\rho$ for different time instants $t$ and for all values of $y$. A nice collapse 
of the data is observed. The diffusion constant $D(\rho)$ starting from zero at $\rho=0$ 
   and  monotonically increases  with $\rho$. 
   
      The high-density limit $\rho \to 1$ is of some interest. In this case, in the steady state, the 
   density of holes is small, the diffusion of holes on a single column looks like the simple exclusion 
   process of particles on a line at low densities. However, the motion of holes in adjacent columns  
   is strongly correlated. For example, if there is a hole at the site $(x,y)$, then a hole can jump 
   to the site $(x, y+2)$ from above only if the positions $(x \pm 1, y+1)$ are occupied by holes.      
   Calculation of limiting value of $D(\rho)$, as $\rho \to 1$ from first principles seems to be an 
   interesting problem.
 
\subsection{ Testing the hydrodynamic equation}

Eq. (\ref{eq:fick_law}) is a rather plausible equation which is an analogue of the Fick's law of diffusion applied to an assembly of strings, which are extended objects. Since the dynamics of strings is much more constrained than the dynamics of point objects, it is useful to test this directly.

      Consider a more general initial condition than shown in Fig. \ref{figure04}. We consider a system still starting with a step profile, but the initial values are $\rho(y,t=0) = \rho_1$, for $y <0$, and $\rho(y,t) = \rho_2$, for $y >0$. This initial condition is also invariant under the transformation $y \rightarrow by$, where $b$ is some scale factor, and consequently, the solution of this equation $\rho(y,t)$ is  also of a function of the single scaling variable $(y/\sqrt{t})$. However, this scaling function now depends on $\rho_1$ and $\rho_2$. This new scaling function, here denoted by $\tilde{\mathcal{F}}( y/\sqrt{t})$, still satisfies Eq. (\ref{eq:unbiased_diffusion}), but now has the boundary conditions $ \tilde{\mathcal{F}}(u)$ tends to $\rho_1$ for $ u \rightarrow -\infty$, and to $\rho_2$, as $u\rightarrow +\infty$.

\begin{figure}[t]
\includegraphics[width=6.0cm]{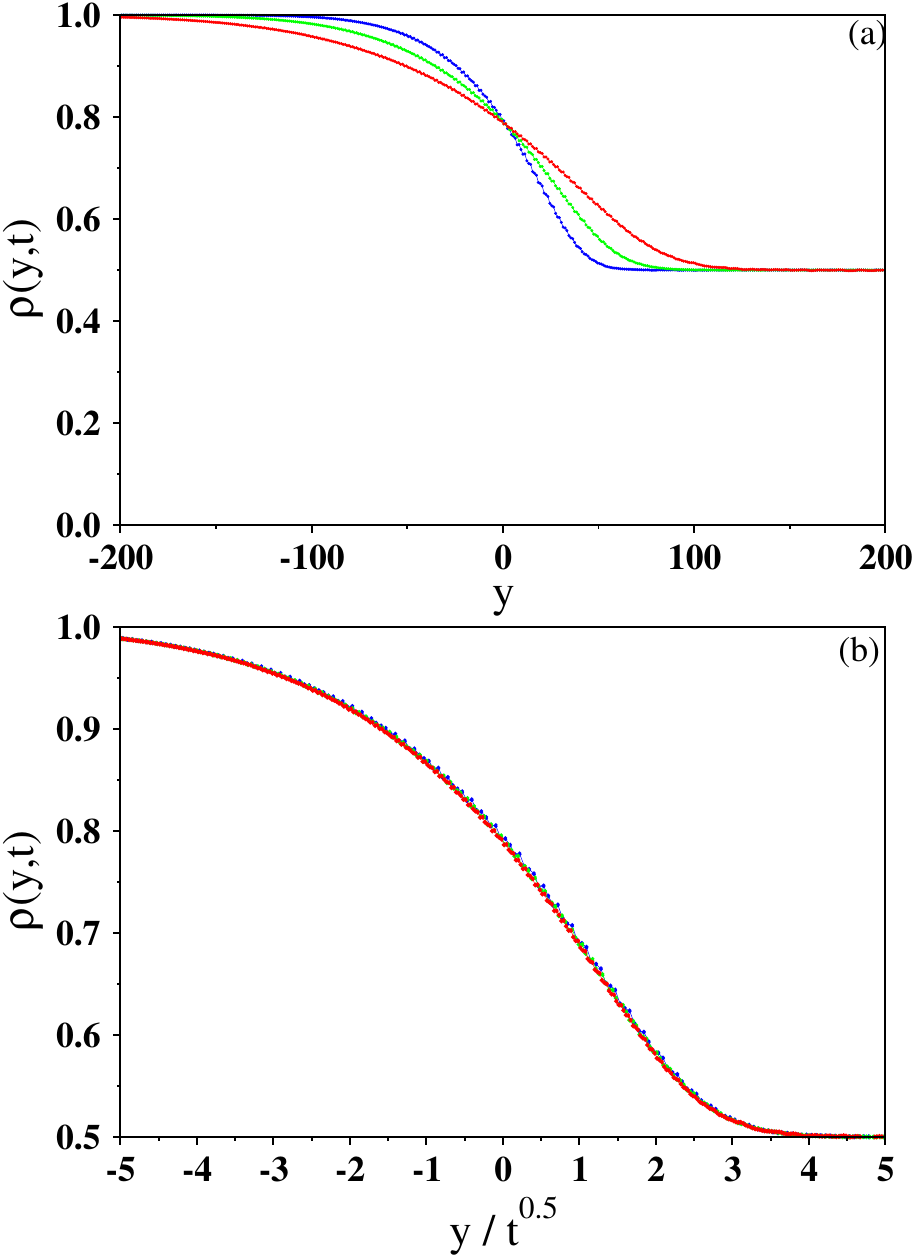}
\caption{Unbiased diffusion for the initial densities $(\rho_1,\rho_2)$=(1,1/2), $L_x$=512.
(a) The average density $\rho(y,t)$  of monomers have been plotted for three different times: 
     $t$ = 256 (blue), 512 (green), and 1024 (red). 
(b) A scaling collapse plot of $\rho(y,t)$ against $y/\sqrt{t}$. }
\label{figure07}
\end{figure}

\begin{figure}[t]
\includegraphics[width=6.0cm]{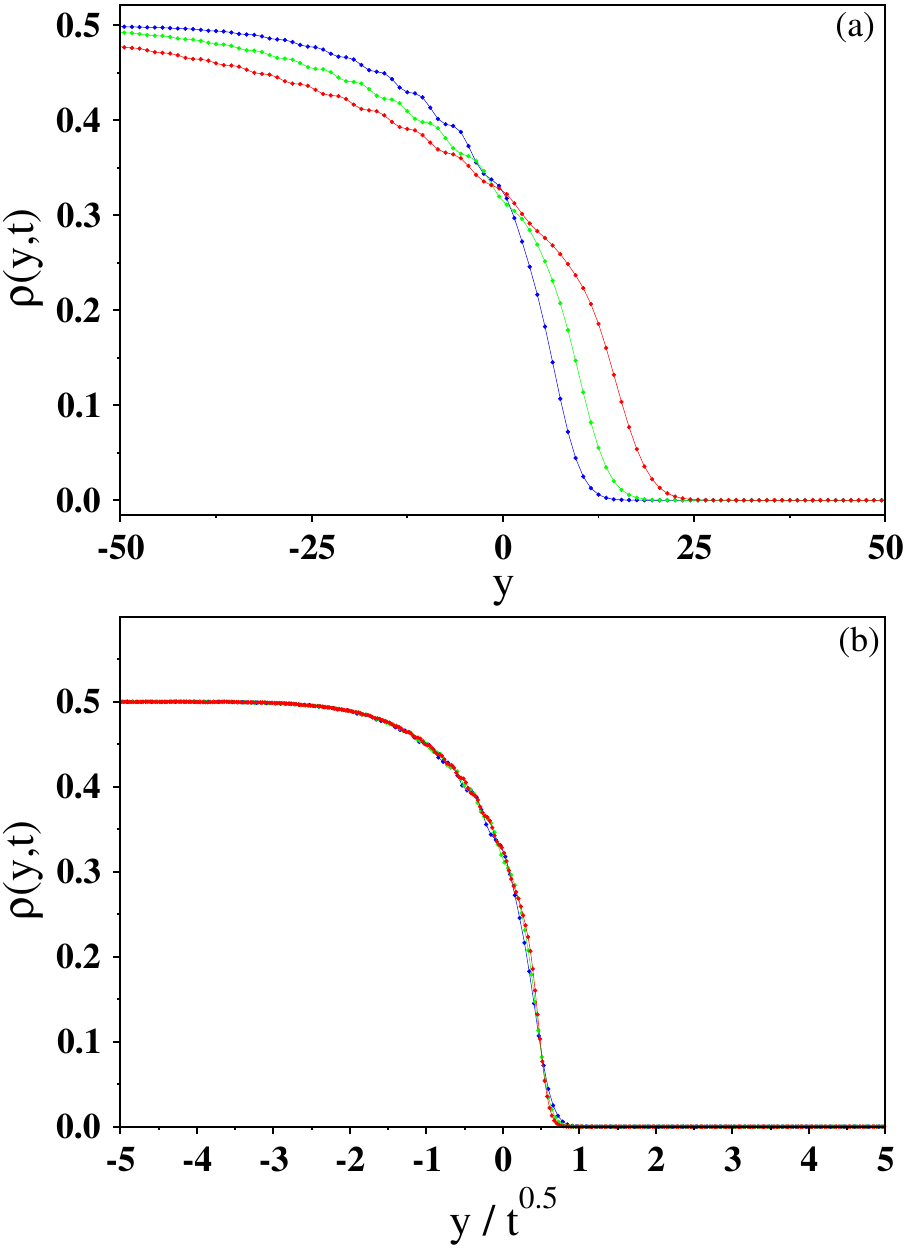}
\caption{Unbiased diffusion for the initial densities $(\rho_1,\rho_2)$=(1/2,0), $L_x$=512.
(a) The average density $\rho(y,t)$  plotted for three different times $t$ :  256 (blue), 512 (green), and 1024 (red).
(b) A scaling collapse plot of $\rho(y,t)$ against $y/\sqrt{t}$. 
}
\label{figure08}
\end{figure}

\begin{figure}[t]
\includegraphics[width=6.0cm]{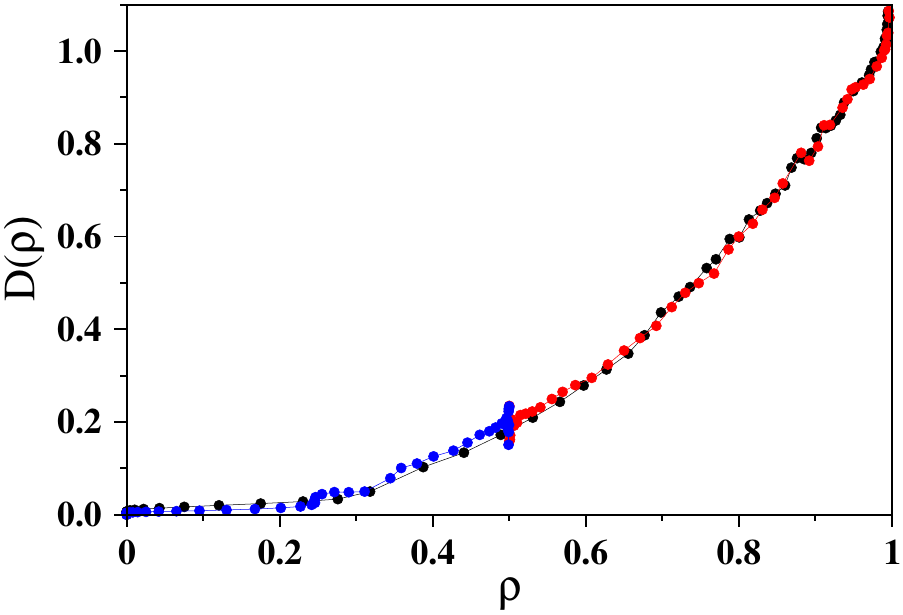}
\caption{Unbiased diffusion on the infinite cylinder of width $L_x$=512.
$D(\rho)$ as a function of  density $\rho$  for three different step boundary conditions: 
$(\rho_1,\rho_2)$=(1,0) (black), (1,1/2) (red), and (1/2,0) (blue).
}
\label{figure09}
\end{figure}

      In Fig. \ref{figure07} we have shown the density profiles at three different times for the initial densities
   $(\rho_1,\rho_2)=(1,1/2)$ and their scaling collapse. Similarly, in Fig. \ref{figure08} the density profiles and 
   the scaling plots for $(\rho_1,\rho_2)=(1/2,0)$ are shown. It is clear that the scaling function in these cases 
   is quite different from that shown in Fig. \ref{figure05} (b). In Fig. \ref{figure09}, we show the function $D(\rho,t)$ 
   plotted against the density $\rho$. We can determine $D(\rho,t)$ only in the density range $1\leq \rho \leq 1/2$ 
   using the the first profile, and in the range $1/2 \leq \rho \leq 0$ using the second profile. But we see that 
   the value of the function $D(\rho,t)$ so determined coincides with the function determined earlier, in these ranges, 
   and  does not depend on the $\rho_1$ and $\rho_2$, to the accuracy of simulations. This is a non-trivial test 
   of the validity of our hydrodynamical equation  Eq. (\ref{eq:unbiased_diffusion}). 


\subsection{Difficulties in the noiseless hydrodynamic description at low densities}
 
      We note that the diffusion coefficient $D(\rho)$ tending zero in the zero-density limit as $\rho^2$ 
   makes the deterministic  hydrodynamic  limit somewhat unrealistic. The diffusion constant becoming zero  
   is valid only in the limit of infinite $L_x$. For finite $L_x$, the center of mass of a single string 
   will show finite diffusion with the diffusion constant $1/L_x$.
 
      If we take the  diffusion equation Eq. (\ref{eq:unscaled_diffusion_equation}) as defining the continuum 
   evolution of a real-valued field $\rho(y,t)$, that is continuous and differentiable  at $t=0$, and the 
   density field is zero at any value $Y$ at some time $T$, then the current through that point is zero, and 
   the diffusing material can never cross that point. If at some time $t$, we have for some value $Y$, the 
   density field tends to zero from the left side, say as  $\rho(Y-\Delta,t) \approx A \Delta$, then the 
   current at $y=Y -\Delta$ is proportional to the diffusion constant, which varies as $\rho^2$, hence as 
   $\Delta^2$. Then, for the particles at this point to up to $y= Y -\Delta/2$ requires a time $T$ of order 
   $\Delta^2/D(\rho)$, which is independent of $\Delta$. Then, the particle will again take the time of order 
   $T$ to reach from $y =Y -\Delta/2$ to  $y=Y-\Delta/4$, and so on, and the time to reach $y=Y$ is divergent. 

      Our Monte Carlo data is consistent with this observation. We find that the scaling function $\mathcal{F}(u)$ 
   does not have a significant tail, and is very nearly zero, for all $ u \gtrsim 1.5$.
    
      In other studies of fronts of non-linear diffusion equations, it is well appreciated that effective 
   diffusion constant at small densities depends on the coarse-graining scale, and the deviations from the 
   noiseless continuum description become significant when the density is so low that the number of particles 
   in the coarse-graining volume is of order one.  See, for example, the discussion in \cite{brunet}. \\
    ~\\

\section{Free fermion description of the steady state in the unbiased case} 

\begin{figure}
\includegraphics[width=4.0cm]{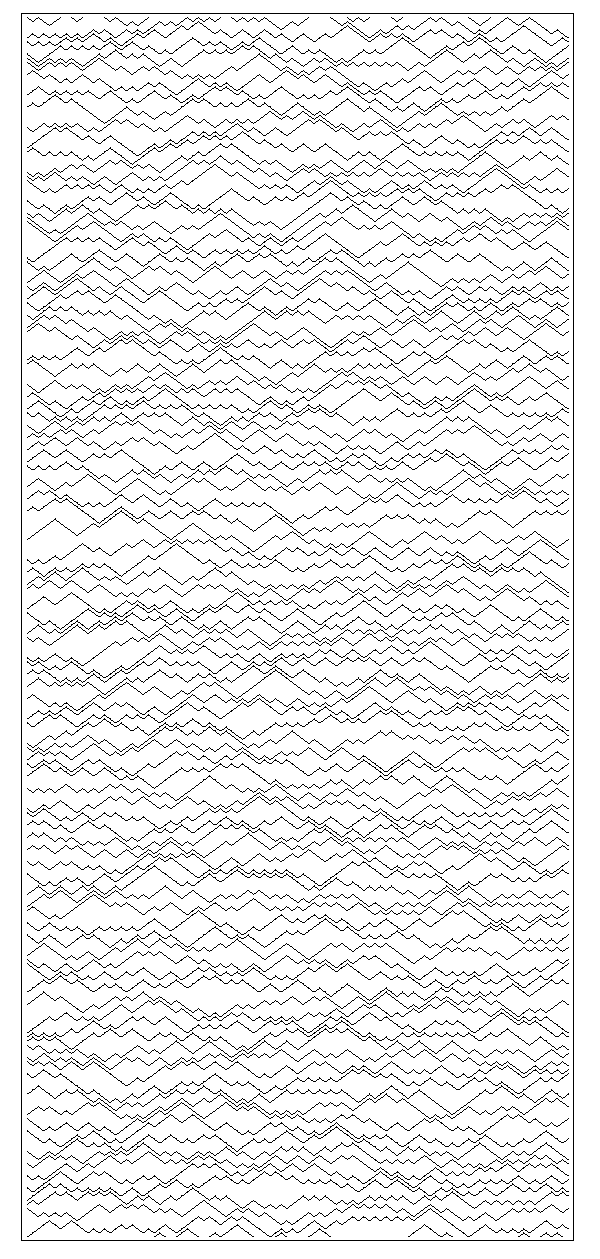}
\caption{Unbiased diffusion:
A typical configuration of the equilibrium steady state of $N$ = 100 unbiased  strings on a torus of size 
$L_x = 100$ and $L_y = 300$ after $t = 15000$ time steps.
}
\label {figure10}
\end{figure}

The simplest setting for studying  the steady state is to consider $N$ strings on a torus, and allow the system to equilibriate.
Here the detailed balance condition is satisfied, and all rates of transition are equal, and hence, in the steady state all allowed configurations of the strings occur with equal probability. An example of a snapshot in the steady state of such a system is shown in Fig. \ref{figure10}. 

Let us denote the total number of allowed configurations, for specified values of $L_x,L_y,N$ by $\Omega(L_x, L_y,N)$. For large systems, this would be expected to vary as $ \exp({L_x N s(\rho)})$, where $s(\rho)$ is called the entropy per monomer, and depends on the density. To be precise, we define
\begin{equation}
s(\rho) = \lim_{ L_x \to \infty, L_y \to \infty} \frac{ \log \Omega(L_x, L_y, N= \rho L_y/2)}{ L_x N}.
\end{equation}

To calculate $s(\rho)$, it is convenient to  consider now the steady state of a set on $N$ fully stretched strings on a rectangular strip of length $L_x$, and height $M$ [Fig. \ref{figure11}]. The ends of the strings may be assumed to be fixed.  The answer is expected to be independent of these boundary conditions in the large system size limit. 

      To determine the number of allowed configurations, we use the transfer matrix approach.  The configuration at $x$ is specified by 
   giving the coordinates $Y_i(x)$, of the $i$-th string for all $i$. Let $Z_x(\{Y\})$ denote the sum over all (partial) 
   configurations of the strings, lying between the lines $x=0$, and  $x$, with the endpoints at $x$ being $\{Y\}  \equiv (Y_1, Y_2, ...Y_N)$ 
   [Fig. \ref{figure01}]. Then, using the transfer matrix formalism, we can express the restricted partition functions 
   vector $Z_{x+1}(\{Y\})$ in terms of the vector $Z_x(\{Y\})$. 
   
 This transfer matrix turns out be rather straightforward, and becomes equivalent to that of free fermions in one dimension. This was first clearly discussed by de Gennes \cite{de_gennes_free_fermions}. We recap it here briefly.
 
 The rules for the transfer matrix are simple. If the configuration at coordinate $x$ is $\{Y\}$ and at $(x+1)$ it is $\{Y'\}$, then we must have $|Y_i' -Y_i| = 1$, and $Y_i' \neq Y_{i+1}'$ for all $i$. In this case, the matrix element is $1$, else $0$. 
 
 It is easy to verify that the eigenfunctions $\psi(\{Y\})$  are of the Slater determinant form made of single particle basis functions $\phi_r(Y_i)$.
\begin{equation}
\psi(\{Y\}) = Det \begin{vmatrix} \phi_1(Y_1)& \phi_1(Y_2)& ...&\phi_1(Y_N)\\
\phi_2(Y_1) & \phi_2(Y_2) &...&\phi_2(Y_N)\\
..& ..& ..& ..\\
\phi_N(Y_1) & \phi_N(Y_2)& ...& \phi_N(Y_N)
\end{vmatrix},
\end{equation}
where the single particle basis functions, for hard wall boundary conditions at $y=0$ and $y=M+1$, are of the form
\begin{equation}
\phi_s(y) = A  \sin k_s y,
\end{equation}
and $k_s(M+1) = m_s \pi$, and $m_s$ are distinct integers. The corresponding eigenvalue $\Lambda$ is given by 
\begin{equation}
\log \Lambda = \sum_{s=1}^{N} \log ( 2 \cos k_s).
\end{equation}

To get the largest eigenvalue, we choose the values of $k_s$ with the largest cosine, i.e. smallest $k_i$. 
Hence we would take $k_s = (\pi s)/(M+1)$. Thus we get the largest value $\Lambda_{max}(N,M)$ given by (take $N$ even, for simplicity)
\begin{equation}
\Lambda_{max}(N,M) = \left[ \prod_{s=1}^{N/2} \left(2\cos(\frac{\pi s}{M+1})\right)\right]^2.
\end{equation}

In the limit of large $M$ and $N$, with $N/M = 2 \rho$ fixed, the sum over k can be converted into integrals, and we get the entropy per monomer $s(\rho)$ in the thermodynamic limit given by
\begin{equation}
s(\rho) = \left(\frac{1}{\rho}\right)\int_{0}^{ \rho \pi} \frac{dk}{  \pi} \left[ \log ( 2 \cos k)\right].
\end{equation}
In the small $\rho$ limit this tends to $ \log 2$, as expected. 

In the steady state of this system, we ask what is the probability that the strings will intersect the vertical line  $x =x_0$  at $\{Y\}$, for some   randomly chosen $x_0$ . Let us denote the eigenvector corresponding to the largest eigenvalue by The eigenvector with the largest eigenvalue was also calculated. Let us denote it by $\Psi_0(\{Y\})$. Then the probability calculation involves a calculation of the form \\

\begin{eqnarray}
Prob(\{Y\}) = \frac{\langle \psi_L|T^{L-x} P_{\{Y\}} T^{x} |\psi_R\rangle} 
	             {\langle \psi_L|T^{L}|\psi_R\rangle}
\nonumber
\end{eqnarray}
Here $\langle \Psi_L|$ and $|\Psi_R\rangle$  are the left and right vectors corresponding to the left and right boundary conditions respectively, and $P_{\{Y\}}$ is the projection operator for the configuration $\{Y\}$. Then, using the fact that if in the stochastic evolution, the system forgets the initial condition to go to a steady state, for $x$ far away from the boundary, \\
$\langle \Psi_L| T^{L_x-x} \approx <\Psi_0| \Lambda^{L_x -x}$,  and $T^x|\Psi_R\rangle \approx \Lambda^{x}|\Psi_0\rangle,$\\
so that we have 
\begin{equation}
{\rm Prob}(\{Y\}) = |\Psi_0(\{Y\})|^2.
\end{equation}
Thus, $\Psi_0(\{Y\})$ is the eigenvector of the transfer matrix with the largest eigenvalue, and acts like the "quantum mechanical wave function" of the ground state of the system of fermions. The fact that the marginal  probabilities of configurations are given by the squared modulus of the quantum mechanical  wavefunction  in this  fully classical problem, has been noted earlier. See, for example, \cite{goldbart1, goldbart2}. 

We note in passing that here this result is obtained in the standard probability theory setting, where the terms in the partition  sum are proportional to {\it the conditional probabilities} of configurations. In quantum mechanics, the terms in the sum of histories are complex amplitudes, whose squared modulus give probabilities. In this case here, we can think of probability ${\rm Prob}(\{Y\})$ of configuration $\{Y\}$, which equals ${\langle \Psi_0|Y\rangle \langle Y|\Psi_0 \rangle}$ as a product of two factors: The ( positive real) factor $\langle Y|\Psi_0\rangle$ as the conditional probability that $Y$ is reached from the initial condition in the past,  and the  (again positive real) factor $\langle \Psi_0| Y\rangle$ is  the conditional probability that $Y$ will reach the prefixed final configuration specified by the boundary conditions at the remote future.  

We would like to mention also the work of Allegra et al, who studied the evolution of a  free fermion gas on a line, with an initial step profile of density \cite{allegra}, but their results are qualitatively different, with spatially inhomogeneous profiles, because of the different boundary conditions used. The related problem of non-intersecting brownian particles on  a line has been studied  in literature under the name of determinantal processes \cite{katori}.
 
\subsection{ A heuristic argument for   $D(\rho) \sim \rho^2$ for small $\rho$:}

      We give  a heuristic  argument  here that $D(\rho) \sim \rho^2$, as $\rho \rightarrow 0$. Consider 
   first the case when there are $N$ nearly horizontal strings, with the average height difference between 
   the $r$-th and $(r+1)$-th string being $\Delta_r$. Here we consider that $\Delta_r \gg 1$, and can vary 
   slowly with $r$.  

      We use the fact that number of walks of $L_x$ steps which are confined between an interval of length 
   $\Delta$ on a one-dimensional line varies as $\lambda(\Delta)^N$, where $\lambda(\Delta) = 2 \cos \frac{\pi}{\Delta+1}$.  
   Then, for large $\Delta$, 
   \begin{equation}
   \log \lambda(L) \approx \log 2 - a/\Delta^2.
   \end{equation}

Then, the entropy of a system per unit length with given values of $\{\Delta_r\}$  will be of the form
$$S(\{\Delta_r\})  \simeq  {\rm Constant } - \sum_{r} \frac {a} { (\Delta_r + \Delta _{r+1})^2}.$$

Then, the thermodynamic pressure  $P$, i.e. the force on the bottom wall of  the  container varies as $ \frac{\partial S}{\partial \Delta_1}  \sim \rho^3$, for small $\rho$. Then, if the pressure varies slowly with $y$, the force on a polymer per unit length is proportional to $\frac{\partial P}{\partial y}$.  In the  the average induced diffusion current would be proportional to $\rho^2 \frac{\partial \rho}{\partial y}$. This implies that $D(\rho) \sim \rho^2 $ for $\rho $ tending to zero. 
    
  We note that the diffusion coefficient becomes zero for zero density,  only in the limit of infinite $L_x$. For finite $L_x$, the center of mass of a single string will show finite diffusion with the diffusion constant $1/L_x$. 
 
\section {The biased diffusion}
   
   In the study of steady state of the system on a torus, we use periodic boundary conditions, and study the resulting steady state for different bulk densities. 
 
 \subsection{Characterizing the biased steady state}
 
    In the biased case, we do not have an exact characterization of the steady state. In the case of totally asymmetric simple exclusion process (TASEP) involving  point particles, while the bias breaks the  detailed balance condition,  for each configuration $C$  of TASEP, the number of configurations from which it can be reached in one step of a single particle always equals the number of configurations to which it can go in one step. As a result, all allowed  configurations are equally likely in the steady state of TASEP \cite{pairwise}.
But,  even this pairwise balance condition does not hold in case of biased  strings. As an example, see the configuration in the Fig. \ref{figure01}, where the number of non-flippable up-kinks is not equal to the number of unflippable down kinks. 
  
\begin{figure}
\includegraphics[width=6.0cm]{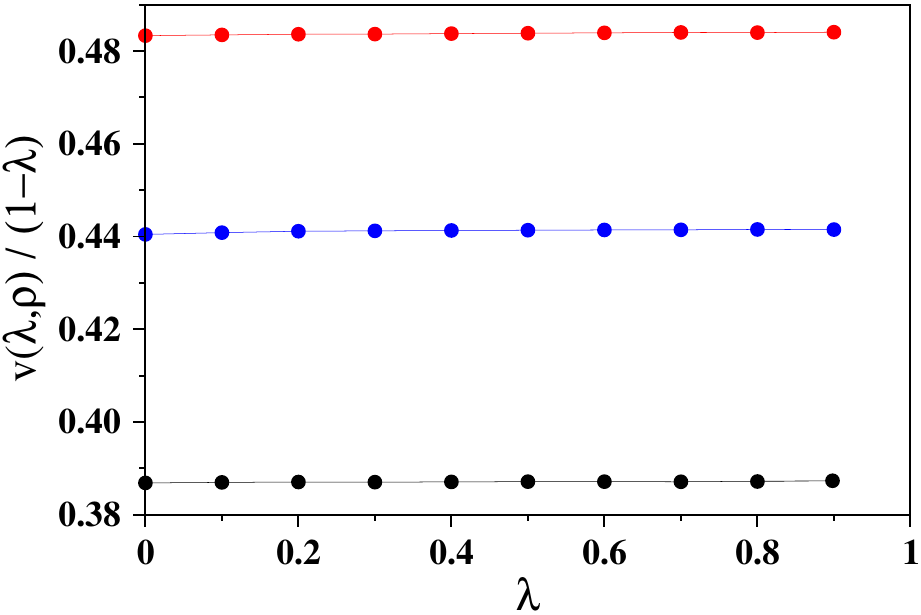}
\caption{Biased diffusion: Plot of the mean velocity $v(\lambda, \rho)$ divided by the factor $(1 -\lambda)$ 
	 as a function of the bias parameter $\lambda$ on a torus of width $L_x$ = 128 and $L_y$ =256, for 
	 three different densities : 0.750 (black), 0.875 (blue), and 0.9375 (red).}
\label{figure11}
\end{figure}

  A bias in the system   produces a steady state in which the center of mass of any string has  a non-zero mean  upward velocity. Let us denote this velocity by $v(\lambda, \rho)$.   We are not able to determine this function from first principles, purely theoretically at present, and will take it as  
   a function  determined  experimentally from simulations. In Fig. \ref{figure11}, we show the mean  drift $y$-velocity of  strings on torus, for  with $L_x$ = 128, $L_y$ = 256, and densities = 15/16, 7/8, 3/4, for ten different values 
   of $\lambda$. If the steady state was independent of $\lambda$, then the density of active upward kinks and downward kinks 
   would be independent of $\lambda$, and the velocity would be proportional to $(1-\lambda)$. The observed values of $v(\lambda, \rho)/((1-\lambda)$ are nearly independent of $\lambda$.  Whether the asymptotic values are exactly proportional to $(1 -\lambda)$ or not, in spite of non-trivial weights of  configurations in the steady state is a very interesting question, that we are not able to address here.

\begin{figure}[t]
\begin {center}
\begin {tabular}{cc}
\includegraphics[width=4.0cm]{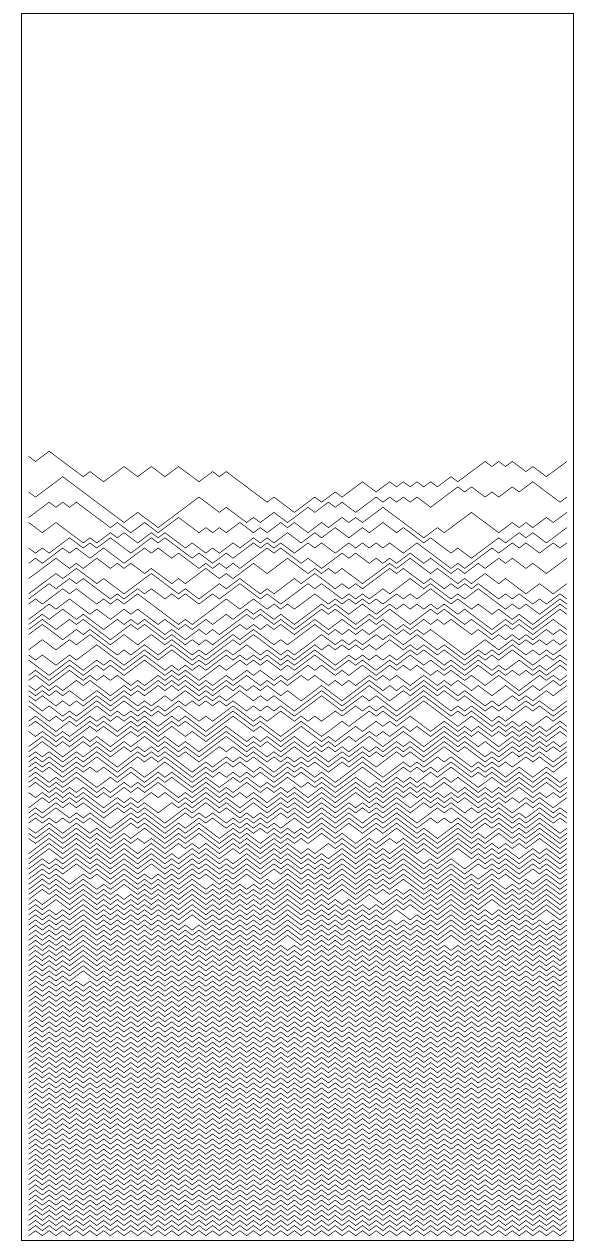} & \includegraphics[width=4.0cm]{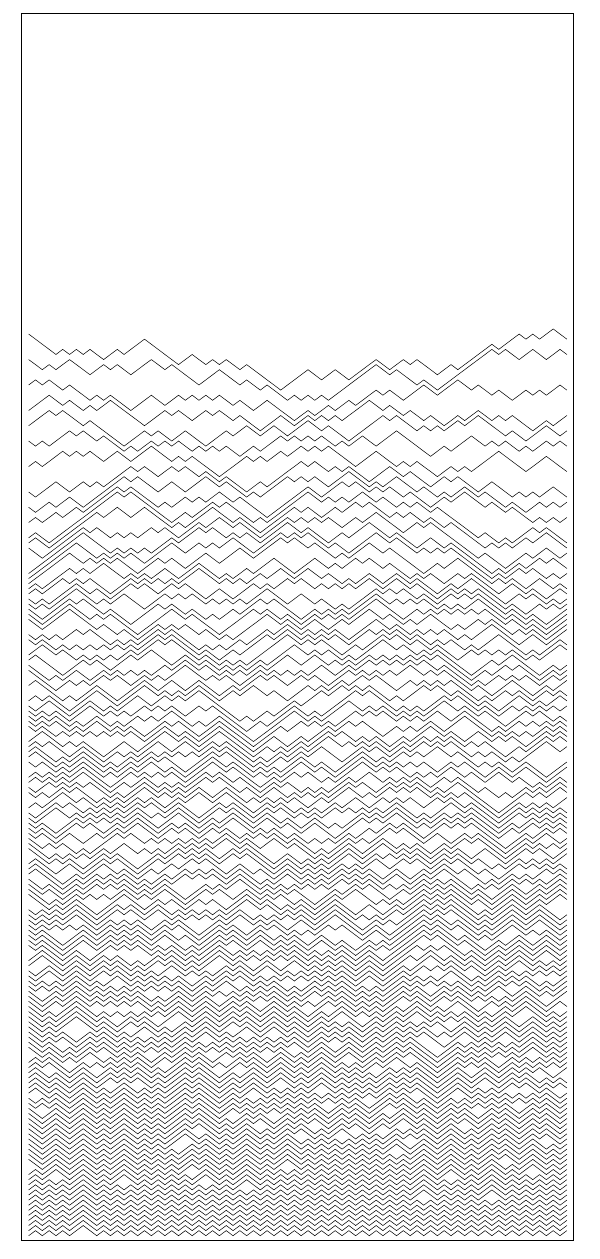}
\end{tabular}
\end {center}
\caption{Biased diffusion:
      Two snapshots of a system of biased strings initially kept with density $1$ for $y<0$ and empty for 
$y>0$, on an infinite cylinder of width $L_x = 80$ after the system has been evolved up to the time $t$ = 
50 and 100 units.
}
\label {figure12}
\end{figure}

\begin{figure}
\includegraphics[width=8cm]{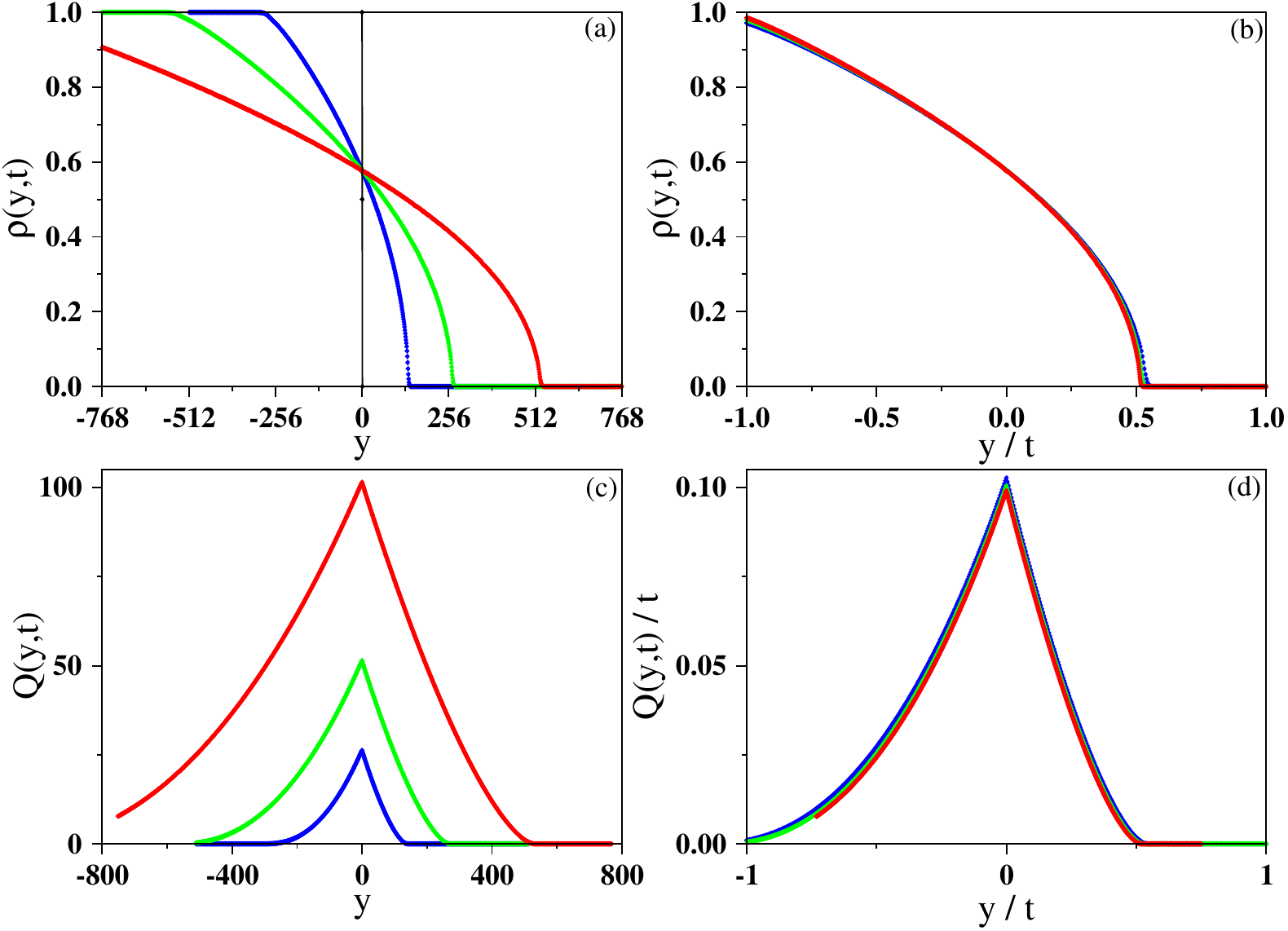}
\caption{(a) The time dependent average density $\rho(y,t)$  of monomers  an half-infinite stack of dense  
   strings on an infinite cylinder of width $L_x$ = 128, plotted at time instants $t$
   = 0 (black), 512 (blue), 1024 (green), and 2048 (red).
   (b) A finite-size data collapse of the same data have been shown. 
   (c) The current density $Q(y,t)$ has been plotted against the vertical coordinate $y$ for a system of 
    biased strings on a lattice of lateral width $L_x$ = 256 at time instants: $t$ = 256 (black), 512 
    (blue) and 1024 (red).
(d) The scaling of the current densities. The scaled plot $Q(y,t) / t$ against $y / t$ exhibits an 
    excellent data collapse for all three times $t$.}

\label{figure13}
\end{figure}

\subsection{The step initial condition}
  
  We consider the step initial condition, where all sites with $y \leq 0$ are fully occupied by zig-zag lines of fully stretched strings, and all sites above this, with $ y > 0$ are all empty. For simplicity, we discuss  only the fully biased case where $\lambda=0$.   At large time scales, in the hydrodynamical description, we expect that the system will show local equilibrium behavior, and in a region with roughly uniform density $\rho$, the system will approximate the behavior  seen in the  steady state on a large torus, with density $\rho$.  Then, we expect that  there will be a mean drift velocity of the strings in the $y$ -direction. Let us denote this by $v^*(\rho)$.

 Then, the equation of continuity implies that $\rho(y,t)$ evolves according to the equation
 \begin{equation}
 \frac{\partial}{\partial t} \rho(y,t) +  \frac{\partial}{\partial y} \left[ v^*(\rho) \rho \right]. 
 \label{eq:biased_hydro}
 \end{equation}
 
 This equation has the scaling symmetry under $ y  \rightarrow a y$ and $t \rightarrow a t$. since the stepped initial condition  also has this symmetry, the evolved profile should also have this symmetry. This implies that the solution  from the stepped profile $\rho(y,t)$ will be only be a function of single  scaling variable  $y/t$. Writing $y/t=u$, we get 
 \begin{equation}
 \rho(y,t) = g(u). 
 \end{equation}
 
      In Fig. \ref{figure13}(a), we have plotted the density profile $\rho(y,t)$ in the fully biased case, 
   for different times. In Fig. \ref{figure13}(b), we show the scaling collapse when $\rho(y,t)$ is plotted 
   against $y / t$. In Fig.\ref{figure13}(c), we show the integrated current density $Q(y,t)$ versus $y$ for 
   three different times $t$. In Fig. \ref{figure13}(d), we show the scaling collapse when $Q(y,t)/t$ is 
   plotted against $y / t$.

      In Eq. \ref{eq:biased_hydro}, writing $\rho(y,t) = g(y/t)$, the scaling function $g(u)$ satisfies the total 
   differential equation
 \begin{equation}
\left(\frac{dg}{du}\right) \left[ -u + \frac{d}{dg} \left[ v^*(g(u)) g(u)\right] \right]=0.
 \end{equation}
 The solution of this equation is quite straight forward: Either $dg/du =0$, and $ g(u)$ is constant, independent of $u$, or we must have 
 \begin{equation}
 u = \frac{d}{dg} \left[ v^*(g) g \right]. 
 \label{eq:scaling_g(u)}
 \end{equation}
 Note that this is no longer a differential equation. Since the functional dependence of $v^*(g) $ is assumed given, it expresses $u$ as explicit  function of $g$. 
 
      In general $v^*(\rho)$ is a decreasing function of $\rho$, and varies from the value $v^*=1/2$ at $g =0$ 
   and $v^*=0$ at $g=1$ with the simple function. If $v^*(\rho) \sim v_{max} - A \rho^2$, for  a fermi gas at low densities, where $A$ is some constant,  Then, $\rho$ as a function of $u$ has a cusp 
   singularity near $u=1/2$, and varies as $(1/2-u)^{1/2}$ for small positive values of $(1/2-u)$.  Our 
   numerical data indicate that $g(u)$ fits well with the simple function $\sqrt{\frac{1}{3}(1-2u)}$, in 
   the entire range $-1\leq u \leq 1/2.$
  
      More generally, we can consider an initial condition where $\rho(y,t) = \rho_1$ for $y<0$ and $\rho(y,t) = \rho_2$ for $y > 0$, with $\rho_2 < \rho_1$. In this case, there are velocities $u_1$ and $u_2$, such that $g(u) = \rho_1$ for $u < u_1$, and $g(u) =\rho_2$ for $u > u_2$. Note that the function  $g(u)$ in the range $u_1 \leq u \leq u_2$ depends only on the function $v^*(g)$, and is independent of the parameters specifying the boundary conditions $\rho_1$ and $\rho_2$. 
 
\section{ Summary and concluding remarks}

      We have defined and studied the stochastic dynamics of directed strings on a lattice interacting
   solely through an excluded-volume constraint as the simplest example of hydrodynamical description 
   of extended interacting objects. We showed that if we start with a  step initial condition on an 
   infinite lattice, the evolution of the mean density is described by a scaling solution in the variable 
   $y/t^{1/2}$ in the unbiased case, and $y/t$ in the biased case. The  scaling function can be determined 
   in terms of quantities defined for steady states of uniform density systems in the large size limit the 
   bulk diffusion constant $D(\rho)$ and the mean velocity $v^*(\rho)$, which depend on the density $\rho$ 
   in a non-trivial way. We estimated these functions from numerical simulations. A first principles 
   theoretical calculation of these coefficients remains an interesting open problem. 

      For the  the unbiased case, the transverse fluctuations of the strings can be exactly described as the 
   quantum mechanical evolution of hard core particles on a line, with the longitudinal direction treated as 
   a kind ``quantum mechanical time''. Then the probabilities of different configurations in the steady state 
   of the system are exactly determined in terms of the eigenfunction of the largest eigenvalue of the transfer matrix. 

      We presented heuristic arguments that the bulk diffusion constant $D(\rho)$ varies as $\rho^2$ for 
   small $\rho$. Exact calculation of $D(\rho)$ and $v(\lambda,\rho)$ remain interesting open problems. 
   Also, one can study fully stretched strings that have an average non-zero slope, which could give rise 
   to a ballistic propagation of the  transverse fluctuations even in the absence of a bias. We have 
   studied in this paper relaxation of density fluctuations in the transverse direction. One can also 
   study the relaxation of localized height-fluctuations of strings, which will also diffuse longitudinally 
   along the string, with a finite diffusion constant, also density-dependent. 

\acknowledgements DD's work is supported by the grant SP/DP/2023/658 by Indian National Science Academy. DD 
   acknowledges very useful discussions with Rajaram Nityananda, and comments on the earlier draft by M. Barma 
   and  H. Spohn. We have used ChatGPT to suggest stylistic changes and additional references on an earlier draft.

\end {document}